\documentclass[12pt]{article}
\font\titlefont=cmbx10 scaled \magstep3
\begin{document}

\begin{center}
{\titlefont What does Quantum Field Theory in Curved Spacetime Have to
Say about the Dark Energy? }

\vskip .3in
L.H. Ford\footnote{email: ford@cosmos.phy.tufts.edu} \\
\vskip .1in
Institute of Cosmology,
Department of Physics and Astronomy\\
Tufts University\\
Medford, Massachusetts 02155\\
\end{center}

\vskip .2in
\begin{abstract}
The issue of the vacuum energy of quantum fields is briefly reviewed. It
is argued that this energy is normally either much too large or much too
small to account for the dark energy, However, there are a few proposals
in which it would be of the order needed to effect the dynamics of the
present day universe. Backreaction models are reviewed, and the question of
whether quantum effects can react against a cosmological constant is discussed.
\end{abstract}

\baselineskip=14pt

\section{Introduction}

The subject of quantum field theory in curved spacetime has been extensively
developed over the past thirty years. These developments include an
understanding of cosmological and black hole particle creation, and a
better understanding of the quantum stress tensor operator. Thus it is
appropriate to ask if any of these insights help to understand the nature
of the dark energy. One can phrase the question as follows:
Is the dark energy of quantum origin, and if so, can it be understood without
radical new physics? 

\section{The Expectation Value of the Quantum Stress Tensor}

If quantum effects are to influence the large scale evolution of the universe,
the simplest description is in terms of a semiclassical theory where 
the expectation value of the stress tensor, $\langle T_{\mu\nu} \rangle$,
acts as the source of gravity. This quantity is formally infinite, and
hence needs to be modified before it can make physical sense. The simplest
modification is a noncovariant frequency cutoff. For a massless field,
such as the electromagnetic field, the energy density  might then become
\begin{equation}
\rho = \frac{1}{(2 \pi)^3}\, \int d^3k\, \omega\, {\rm e}^{-\alpha \omega} \,,
                                              \label{eq:noncov}
\end{equation}
where $\alpha$ is an arbitrary cutoff parameter
and the pressure is $P = \frac{1}{3} \rho$. The obvious objection to this
modification is that it violates Lorentz invariance. However, in the context
of cosmology, there is a preferred frame so this cutoff may not be quite
as crazy as it seems. The energy density can be written as a function of
$\alpha$ as
\begin{equation}
\rho = 10^{-30} {\rm \frac{gm}{cm^3} } 
       \left(\frac{10^{-2} {\rm cm}}{\alpha} \right)^4 \,.
\end{equation}
There are several problems with this result. There would not seem to be any
natural reason to select $\alpha \approx 10^{-2} {\rm cm}$ as a cutoff. A
much smaller value of $\alpha$ leads to a radiation dominated universe which
recollapses in less than the age of the present universe. In any case, this
model does not describe anything resembling the dark energy.

A better approach is to use a covariant cutoff, which preserves local
Lorentz invariance. There are various covariant regularization methods
which have been developed, including dimensional regularization,
zeta-function regularization, and covariant point-splitting with
direction averaging \cite{BD82}. 
For our purposes, the details of these techniques
are not important. The key result is that the divergent parts of
$\langle T_{\mu\nu} \rangle$ may be written in terms of geometrical
quantities as follows:
\begin{equation}
\langle T_{\mu\nu} \rangle_{\rm div} = A\:{{g_{\mu\nu}}\over \beta^4} + 
B\:{{G_{\mu\nu}}\over \beta^2} + \bigl(C_1 H^{(1)}_{\mu\nu} +
C_2 H^{(2)}_{\mu\nu} \bigr)\: \ln\,\beta. \label{eq:cov}
\end{equation}
Here $\beta$ is a cutoff parameter with the dimensions of length,
$A$, $B$, $C_1$, and  $C_2$ are constants, $G_{\mu\nu}$ is the Einstein 
tensor, and the $H^{(1)}_{\mu\nu}$ and $H^{(2)}_{\mu\nu}$ tensors are 
covariantly conserved tensors which are quadratic in the Riemann tensor.
Specifically, they are the functional derivatives with respect to the metric
tensor of the square of the scalar curvature and of the Ricci tensor, 
respectively:
\begin{eqnarray}
H^{(1)}_{\mu\nu} &\equiv& {1\over \sqrt{-g}} {\delta \over {\delta g^{\mu\nu}}} 
                                 \bigl[\sqrt{-g} R^2 \bigr] \nonumber \\
&=& 2\nabla_\nu \nabla_\mu R -2g_{\mu\nu}\nabla_\rho \nabla^\rho R
 - {1\over 2}g_{\mu\nu} R^2 +2R R_{\mu\nu},  \label{eq:H1}
\end{eqnarray}
and
\begin{eqnarray}
H^{(2)}_{\mu\nu} &\equiv& 
                {1\over \sqrt{-g}} {\delta \over {\delta g^{\mu\nu}}} 
            \bigl[\sqrt{-g} R_{\alpha\beta}R^{\alpha\beta} \bigr] 
= 2\nabla_\alpha \nabla_\nu R_\mu^\alpha - \nabla_\rho \nabla^\rho R_{\mu\nu}
\nonumber \\ &{}& -{1\over 2}g_{\mu\nu}\nabla_\rho \nabla^\rho R
  -{1\over 2}g_{\mu\nu} R_{\alpha\beta}R^{\alpha\beta} 
   +2R_\mu^\rho R_{\rho\nu}.    \label{eq:H2}
\end{eqnarray}
Let us focus our attention on the leading term, that proportional to
$\beta^{-4}\,g_{\mu\nu}$. This is of the form of a cosmological constant,
so that the corresponding equation of state is $P = - \rho$. Note that
the price of a covariant regularization scheme is the breaking of conformal
invariance, so massless fields no longer have traceless stress tensors.
If $\beta \ll 10^{-2} {\rm cm}$, then the resulting cosmological constant
is too large to be consistent with the observed universe, and must be
removed by renormalization.

All of the cutoff-dependent terms in $\langle T_{\mu\nu} \rangle$ may be 
absorbed into redefinitions of the constants appearing in the gravitational 
action
\begin{equation}
S_G = {1\over {16\pi G_0}} \int d^4x\,\sqrt{-g}\, \Bigl( R -2\Lambda_0
      + \alpha_0 R^2 + \beta_0 R_{\alpha\beta}R^{\alpha\beta} \Bigr).
\end{equation}
We now include a matter action, $S_M$, and vary the total action, 
$S= S_G + S_M$, with respect to the metric. If we replace the classical
stress tensor in the resulting equation by the quantum expectation value,
$\langle T_{\mu\nu} \rangle$, we obtain the semiclassical Einstein equation
including the quadratic counterterms:
\begin{equation}
G_{\mu\nu} +\Lambda_0 g_{\mu\nu} +\alpha_0 H^{(1)}_{\mu\nu}
 +\beta_0 H^{(2)}_{\mu\nu} = 8\pi G_0 \langle T_{\mu\nu} \rangle.
\end{equation}
We may remove the divergent parts of $\langle T_{\mu\nu} \rangle$ in 
redefinitions of the coupling constants $G_0$, $\Lambda_0$, $\alpha_0$, 
and $\beta_0$.
 
However, the {\it renormalized} values of these constants are free parameters
which cannot be calculated by the theory. At this level, quantum field
theory can no more calculate the cosmological constant $\Lambda$ than it can
find Newton's constant or the mass of the electron. Thus, so far the answer
to the question in the title of this article is ``Nothing!''. 

We might next inquire about the finite part of $\langle T_{\mu\nu} \rangle$
which is not of the form of any of the counterterms used in renormalization.
This part is unambiguous, and can be explicitly calculated for simple models,
such as a massless scalar field in a Friedmann-Robertson-Walker 
universe \cite{Bunch}.
Unfortunately, the results are typically of order 
\begin{equation}
\langle T_{\mu\nu} \rangle_{\rm fin} \approx \frac{1}{t^4} \, , \label{eq:t4}
\end{equation}
where $t$ is the present age of the universe. This is too small by a factor
of about $10^{-120}$ to alter the dynamics of the universe at the present
time. Conversely, if we use the cutoff dependent expressions 
Eqs.~(\ref{eq:noncov}) or (\ref{eq:cov}) with $\beta$ of the order of the 
Planck scale, our answer is too large by a factor of about $10^{120}$.
It is not clear how to find a result which is the geometric mean of these
two extremes in a natural way. In other words, a cosmologically interesting
energy density arises from a scale of the order of $10^{-2} {\rm cm}$, which
is about the geometric mean of the size of the observable universe and the
Planck length. It is far from clear why such a length scale should arise.

There have been a number of ideas proposed for mechanisms which might solve 
this puzzle by providing a model for the finite part of 
$\langle T_{\mu\nu} \rangle$ which is much larger than given in 
Eq.~(\ref{eq:t4}). Parker and Raval\cite{PR99} have suggested that there could 
be a large contribution from low mass scalar fields, leading to a limiting 
value of the scalar curvature. In this model, the scalar curvature could
drop as in standard cosmology until this limiting value is approached.
Then the universe would enter a phase of accelerated expansion.
Another model, due to Sahni and Habib\cite{SH98} also postulates a 
low mass scalar field.
In this model, the $\langle T_{\mu\nu} \rangle$ due to created particle
is large and approximately of the form of a cosmological constant term.
Sch\"{u}tzhold\cite{Schutzhold} has proposed a model in which the QCD trace 
anomaly may produce a term in the stress tensor of the form
\begin{equation}
\langle T_{\mu\nu} \rangle_{\rm fin} \approx \frac{\Lambda_{QCD}^3}{t}\, ,
\end{equation}
where $\Lambda_{QCD} \approx 10^2 {\rm MeV}$ is the QCD scale. At the present
age of the universe, $t$, this term has about the right order of magnitude
to begin to dominate the cosmological expansion.
The three models discussed in this paragraph are all somewhat speculative,
but indicate that there are possible ways to get cosmologically
significant energy densities in the present epoch from quantum effects.

\section{Backreaction Models}

Now I wish to turn to a class of models which attempt to explain why 
the cosmological constant term is not enormously large today, and also
possibly explain there may be an effective cosmological constant term
which is large enough to alter the present expansion rate. These are
backreaction models (or adjustment mechanisms). The basic idea is that
some type of instability which causes a large value of $\Lambda_{eff}$
in the early universe to decay naturally to a smaller value today. Ideally,
one would like have a mechanism which act slowly enough to allow inflation
to proceed. Thus if deSitter space is unstable, it should be so on a scale
of more than about sixty horizon lengths, the minimum time needed for
inflationary models to explain the horizon and flatness problems.  One
would also like a natural evolution to 
$\Lambda_{eff} \leq 10^{-30} {\rm g/cm^3}$ today. In such a model, the
effective cosmological constant is now very small compared to particle 
physics energy scales because the universe is very old compared to particle
physics time scales.

No compelling mechanism which accomplishes both of these goals has yet been
found, and doubts have been expressed as to whether such a mechanism
can exist in principle \cite{Weinberg}. Nonetheless, it is worth looking
at some of the possibilities, as a successful backreaction model would be
a great advance. It is also likely that any such model would rely upon
quantum effects. 

\subsection{Quantum Instability of deSitter Space?}

One way in which backreaction against a cosmological term could manifest 
itself is through an instability of deSitter space, the solution which
is the attractor in the set of solutions of Einstein's equations with a
positive cosmological constant. There is one example of a quantum instability
{\it in} deSitter space, if not  {\it of} deSitter space. This is the case
of a free, massless, minimally coupled scalar field, and arises from the 
infrared behavior\cite{FP77a} of this theory. 
A massive scalar field has a well-defined
deSitter invariant vacuum state. However, the two-point function diverges
in the limit that the mass $m$ goes to zero:
\begin{equation}
\langle \varphi(x) \varphi(x') \rangle \sim \frac{1}{m^2}\,  \qquad
 m \to 0 \,.
\end{equation}
There exist a class of quantum states which are free of the infrared
divergence, but which all break deSitter invariance. In all of these states,
$\langle \varphi^2 \rangle$ becomes a function of time. In particular,
in the representation of deSitter space as a spatially flat Robertson-Walker
universe, all infrared-finite states lead to linear 
growth\cite{VF82} in the comoving time, $t$:
\begin{equation}
\langle \varphi^2 \rangle  \sim \frac{H^3 t}{4 \pi^2}\,  \qquad t \to \infty\,,
\end{equation}
where $H$ is the inverse expansion time. 
However, this instability of the massless free field does not not lead to any 
backreaction on the rate of expansion. The reason is that the stress tensor
for this field is bounded
\begin{equation}
\langle T_{\mu\nu} \rangle \to {\rm constant} \rangle \,  \qquad t \to \infty\,.
\end{equation}
The derivatives of $\varphi$ in the expression for $ T_{\mu\nu}$ remove
the contribution of the long wavelength modes which cause 
$\langle \varphi^2 \rangle$ to grow.

In the case of an interacting field theory involving a massless scalar field,
it is possible for $\langle T_{\mu\nu} \rangle$ to grow for a finite amount 
of time\cite{F85,AW01}. 
Consider, for example, a self-coupled field with a $\lambda \varphi^4$
interaction. In this case, there will be a term in the stress tensor
of the form of $\lambda  \langle \varphi^2 \rangle^2 g_{\mu\nu}$, which will
initially grow as $t^2$. This growth will only last a finite time, however,
before higher order contributions become important and act to stop the
growth of $\langle T_{\mu\nu} \rangle$. The field begins to acquire an
effective mass, which suppresses the infrared instability driving the
growth.

A more promising source of instability comes when we quantize the gravitational
field on the deSitter background. Consider classical gravitational wave
perturbations of a spatially flat  Robertson-Walker universe. If we impose 
the transverse-tracefree gauge condition, which removes all gauge freedom
and is the analog of the Coulomb gauge in electrodynamics, then the 
independent components of the perturbation are equivalent to a pair of
massless scalar fields\cite{Lifshitz,FP77b}. 
This means that linearized quantum gravity on a 
deSitter background is subject to the same infrared instability as is the
massless scalar field. In particular, the mean square of the perturbation
will grow in time:
\begin{equation}
\langle h_{\mu\nu} h^{\mu\nu} \rangle \sim  t\,  \qquad t \to \infty\,.
\end{equation}
However, at this one loop level, there is no backreaction on the expansion,
just as for the case of scalar fields. This raises the question of what
happens in the next, two loop, order. This question has been examined by
Dolgov, {\it et al} \cite{DEZ} and especially by
Tsamis and Woodard \cite{TW}, who find that the effective stress tensor
of the gravitons does grow in this order, and will tend to react against
the deSitter expansion. Unfortunately, the backreaction only begins to
be significant at the point that the two-loop approximation becomes
questionable, and higher loop corrections cannot be ignored. This leaves us
with the intriguing possibility that quantum gravity may be a source of
backreaction, but no means of testing this possibility beyond two loop
perturbative quantum gravity.

\subsection{Dolgov-type Models}
\subsubsection{The Original Dolgov Model}

In 1982, Dolgov \cite{Dolgov} proposed a remarkably simple classical model
for backreaction, based upon a massless, nonminimally coupled scalar field.
This field has the Lagrangian
\begin{equation}
{\cal L} = \frac{1}{2} ( \partial_\alpha \varphi \partial^\alpha \varphi
- \xi R \varphi^2) \, ,   \label{eq:Lagrangian}
\end{equation}
where $R$ is the scalar curvature and $\xi$ is a {\it negative} constant.
The associated equation for $\varphi$ is
\begin{equation}
\nabla_\alpha \nabla^\alpha \varphi + \xi R \varphi = 0 \, ,
\end{equation}
and has growing solutions if $\xi < 0$ and $R > 0$. In deSitter space,
where $R$ is a positive constant, the unstable solution grows exponentially:
\begin{equation}
\varphi(t) \sim e^{\gamma t} \,,
\end{equation}
where 
\begin{equation}
\gamma = \frac{3}{2} H \left[ \left(1+
        \frac{16}{3} |\xi| \right)^\frac{1}{2} -1 \right] \,.
\end{equation}
Here $t$ is the comoving time in the spatially flat  Robertson-Walker 
coordinates. The stress tensor of this mode also grows exponentially,
\begin{equation}
\langle T_{\mu\nu} \rangle \sim e^{2\gamma t} \,,
\end{equation}
and causes the system to exit deSitter space on a time scale of the order
of $1/\gamma$. This time scale is in turn determined by the coupling
constant $\xi$, and can be long compared to the expansion time $1/H$
if $|\xi| \ll 1$. 

At late times, the scalar field grows linearly in time
\begin{equation}
\varphi \sim \lambda t\, ,  \label{eq:phi_rate}
\end{equation}
where $\lambda$ is a constant determined by $\xi$, and the scale factor
grows as a power of time
\begin{equation}
a(t) \sim t^\alpha\, \qquad \alpha = \frac{2|\xi| +1}{4|\xi|} \,.
\end{equation}
Most importantly, the scalar field stress tensor approaches the form of
a cosmological constant term:
\begin{equation}
\langle T_{\mu\nu} \rangle \sim \frac{1}{8 \pi}\, \Lambda_0\, g_{\mu\nu}
 + O(t^{-2}) \,,
\end{equation}
where $\Lambda_0 = 3 H^2$ is the effective value of the cosmological constant
during the deSitter phase. The leading term in $\langle T_{\mu\nu} \rangle$
cancels the effect of this cosmological constant. The remarkable feature
of the backreaction is that it provides a natural cancellation of the
original cosmological constant for any value of $\Lambda_0$ and to just
the accuracy needed. The residual term of order $t^{-2}$ is of just the 
magnitude needed to be cosmologically significant at the present. It should
be noted that Dolgov's model evades Weinberg's ``no-go'' theorem 
\cite{Weinberg}, which
attempted to rule out backreaction models. Weinberg's  theorem assumes
that all fields are asymptotically constant in the future, which is not the
case here, as can be seen from Eq.~(\ref{eq:phi_rate}). The Dolgov model
basically uses the kinetic energy of the growing $\varphi$ field to cancel
the cosmological constant.

Unfortunately, this model also suffers from a fatal flaw, which was recognized
in the original paper of Dolgov: The effective value of Newton's constant
is also driven to zero. This arises because the $R$ term in 
Eq.~(\ref{eq:Lagrangian}) is of the form of the Lagrangian for gravity.
This leads to an effective value  of Newton's constant of
\begin{equation}
G_{eff} = \frac{G_0}{1 + 8\pi G_0 |\xi| \varphi^2} \sim \frac{1}{t^2} \,,
\end{equation}
where $G_0$ is the ``bare'' value of  Newton's constant when $\varphi = 0$.
There is a lesser flaw in the Dolgov model as well. In order to achieve adequate 
inflation, we want $|\xi| \ll 1$, which implies that $\alpha \gg 1$. In this 
case, inflation never really ends and we do not have a very realistic cosmology.

\subsubsection{A Variant of the Dolgov Model}

An alternative model \cite{F87} uses the Lagrangian 
\begin{equation}
{\cal L} = \frac{1}{2} \left[ \partial_\alpha \varphi \partial^\alpha \varphi
- \xi_0 \, R\, \ln(R \ell^2)\, \varphi^2 \right] \, ,   \label{eq:Lagrangian2}
\end{equation}
where $\ell$ is an arbitrary length scale.
The motivation for the introduction of the $\ln(R \ell^2)$ factor comes from
quantum effects in curved spacetime. It can be shown from renormalization 
group arguments that $\langle \varphi^2 \rangle$ for a free field always
acquires the term\cite{FT84}
\begin{equation}
\langle \varphi^2 \rangle_{R} = 
\frac{\xi -\frac{1}{6}}{96 \pi^2}\,  R\, \ln(R \ell^2) \,.
\end{equation}
This means that one loop quantum corrections will cause a self coupled
scalar field with a $\lambda \varphi^4$ interaction to acquire a 
 $\ln(R \ell^2)$ term of the form of that in Eq.~(\ref{eq:Lagrangian2}).
The effect of this term is to cause $\xi$ to become a running coupling constant
which scales with curvature. The effective value of $\xi$ is
\begin{equation}
\xi_{eff} = \xi_0 \, \ln(R \ell^2) \,.
\end{equation}
The time dependence of $\xi_{eff}$ is sufficiently weak that we can 
arrange that it has
one constant value during the initial deSitter phase, and another approximately
constant value today. In particular, it is possible to have $|\xi_{eff}| \ll 1$
in the deSitter phase, but $|\xi_{eff}| \approx 1$ today. This solves the
problem in the original model that one could not have both adequate inflation
in the early universe, and noninflationary expansion today.

Unfortunately, the bigger problem of the vanishing of gravity on scales
small compared to the horizon remains in this new model. However, we can see
what might solve this problem. If one had a compelling reason to treat
Eq.~(\ref{eq:Lagrangian2}) only as an effective action for deriving the 
equation for $\varphi$, but not the Einstein equations, then the problem
would disappear. In other words, one needs a model in which quantum corrections
create a Dolgov-type model on cosmological scales, but which do not modify 
gravity on much smaller scales. It is not clear if such a model is possible,
but this seems to be worth exploring.

\section{Summary}

We have seen that the cutoff-dependent vacuum energy of quantum fields,
whether given by Eq.~(\ref{eq:noncov}) or by Eq.~(\ref{eq:cov}),
is much too large if the cutoff is dictated by any particle physics
length scale. If the cutoff is at the Planck scale, then the answer is
too large by a factor of about $10^{120}$. On the other hand, the 
renormalized energy density, such as that in  Eq.~(\ref{eq:t4}), is too
small to influence the present expansion of the universe by a factor
 of about $10^{-120}$. There are several models which give an intermediate
value which is neither too large nor too small to be of interest.
However, all of these models are somewhat speculative. It is not yet clear
that there is a natural and compelling way to get a cosmologically interesting
energy density from quantum effects.

A viable backreaction model would seem to be the best way both to
resolve the fine tuning problem and to explain the dark energy. The 
outstanding question is whether any such model exists. The quantum
instability of gravitons in deSitter space provides a possible route
for the onset on an instability. The difficult problem is describe what
happens next. The Dolgov model gives an intriguing picture of the form 
that late time backreaction might take. It remains to be seen if this
sort of behavior can arise in a realistic theory.

\vspace{0.5cm}

{\bf Acknowledgement:} This work was supported in part by the National Science
Foundation under Grant PHY-9800965.



\vfill
\end{document}